# BLIND AND ROBUST IMAGES WATERMARKING BASED ON WAVELET AND EDGE INSERTION

Henri Bruno Razafindradina[1] and Attoumani Mohamed Karim[2]

[1]Higher Institute of Technology, Diego Suarez - Madagascar
[2]Department of Studies and Projects at the Comoros ICT Regulation Authority, Comoros

*ABSTRACT*

*This paper gives a new scheme of watermarking technique related to insert the mark by adding edge in HH sub-band of the host image after wavelet decomposition. Contrary to most of the watermarking algorithms in wavelet domain, our method is blind and results show that it is robust against the JPEG and GIF compression, histogram and spectrum spreading, noise adding and small rotation. Its robustness against compression is better than others watermarking algorithms reported in the literature. The algorithm is flexible because its capacity or robustness can be improved by modifying some parameters.*

*KEYWORDS*

*Watermarking, Wavelet, Edge, Multimedia, Copyright*

## 1. INTRODUCTION

Watermarking is a technique which consists in inserting a robust and imperceptible brand in a host image, in order to protect it against illegal copying. The watermarking algorithms must be imperceptible to the naked eye, robust against attacks, blind which means : the original image is not necessary for the detection and extraction of the brand. The wavelet decomposition is much used in compression, denoising and image watermarking. It has shown its efficiency in compression with the birth of the JPEG2000 standard. About the use of wavelet watermarking, compared to other techniques such as Fast Fourier Transform (FFT - Fast Fourier Transform), the Discrete Cosine Transform (DCT - Discrete Cosine Transform), Spread Spectrum (CDMA - Code Division Multiple Access), the number of publications in the field of wavelet transform (DWT - Discrete Wavelet Transform) does not cease to increase. Watermarking scheme such as the insertion by adding the wavelet coefficients of the brand in the high and low frequency bands of the image host [1] [2], based on the Delaunay triangulation [3] algorithm, resist several types of attacks. Several attempts have even been made to combine DWT with other transformed, can be cited : the Ganic and Al [4] algorithm that inserts the singular values of the brand in those of the host image, video watermarking proposed by Fan [5] which spreads the bits of the mark before its insertion, Jiansheng [6] method that inserts the DCT of the mark in the HH band of the host image after decomposition and finally other hybrid methods [7] [8] [9] [10] [11] [12] [13] based on mixed transform techniques. Certainly, these techniques are efficient, but most of them are not blind.

In this paper, we present a new robust and blind watermarking method that change the coefficients of the HH band by adding edges. The extraction of the brand consists in detecting the edges in the same band. The basics of the wavelet transform are first described, then we will detail the proposed method and the results will be discussed.

## 2. THE WAVELET TRANSFORM

The Wavelet Transform is a multi-resolution description of an image. It decomposes an image into several sub-bands in three different directions : horizontal, vertical and diagonal. It consists to decompose an image into low and high frequencies using respectively low-pass and high-pass filters [15] :

$$H(\omega) = \sum_k h[k]\, e^{-jk\omega} \quad \text{and} \quad G(\omega) = \sum_k g[k]\, e^{-jk\omega} \tag{1}$$

where H(ω) and G(ω) must be orthogonal :

$$|H(\omega)|^2 + |G(\omega)|^2 = 1 \tag{2}$$

The obtained coefficients are :

$$c[j-1,k] = \sum_n h[n-2k]\, c[j,n]$$
$$d[j-1,k] = \sum_n g[n-2k]\, c[j,n] \tag{3}$$

The reconstruction IDWT of the original signal is the reverse process of DWT. It is summarized by the following formula:

$$c[j,n] = \sum_k h[n-2k]\, c[j-1,k] + \sum_k g[n-2k]\, d[j-1,k] \tag{4}$$

The result is an approximation image having a halved resolution and three detail images which give the errors between the original image and the approximation image. This transformation is repeated as many times as necessary to obtain the desired number of sub-bands. After a few levels of decomposition, the low frequencies are concentrated on the top left corner of the transform and look like a compressed version of the original image.

## 3. PROPOSED WATERMARKING SCHEME

### 3.1. Our Approach

Despite the complexity of the computation of the wavelet coefficients, the DWT is much closer to the human visual system [14] that the DCT or DFT. In the watermarking insertion part, the first stage is to decompose the image into four frequency bands denoted LL, HL, LH and HH.

International Journal on Cryptography and Information Security (IJCIS), Vol.3, No. 3, September 2013

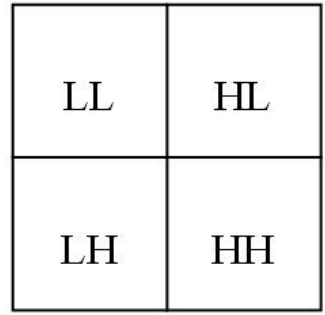

Figure 1. One level decomposition with DWT

The proposed algorithm (second stage) inserts a bit of the brand by adding a block of size N × N in the HH band (high frequency) : the first block (a) shows a bit "1" and the other block ( b) represents a bit "0" or vice versa.

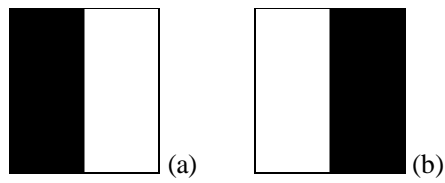

Figure 2. Block representing bits to be inserted

These two blocks (a) and (b) can be modeled as edges : the first is a rising and the second is a falling edge. Therefore, their gradient (formula 5) along the horizontal axis is a peak for (a) and a hollow for (b).

$$GradBloc_x(x,y) = \frac{\partial Bloc(x,y)}{\partial x} \tag{5}$$

The following figure shows the appearance of gradients obtained by differentiating the two blocks in Figure 2.

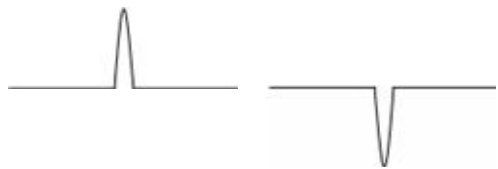

Figure 3. Respective gradients of the rising edges (a) and falling (b)

Adding marks inserts rising or falling edges in the image. The extraction of the mark detects the presence of these edges in the image.

The following figure shows the proposed method :

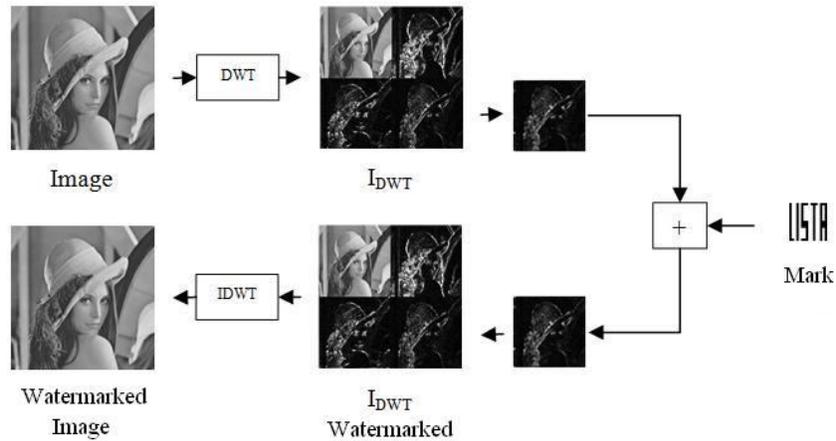

Figure 4. Insertion method

## 3.2. Insertion Algorithm

To insert a bit, selecting a block noted blocHH of the same size than the mark in the HH band and then adding two blocks. The insertion algorithm is summarized as follows :

1. Compute the wavelet transform of the image I (we denote this transform IDWT).
2. Add brand in blocHH block by applying the formula :
    If (Bit_Of_The_Mark = 1) then
        blocHH = blocHH + $\lambda \times$ rising edge
    Else
        blocHH = blocHH + $\lambda \times$ falling edge
    End If
3. Finally, the inverse transformation is applied to the IDWT image to form the watermarked image.
Where $\lambda$ is a factor to adjust the robustness and imperceptibility of the technique. In this algorithm, the edges are matrix formed by -1 (black) and 1 (white).

## 3.3. Extraction Algorithm

The following figure illustrates the extraction of the mark :

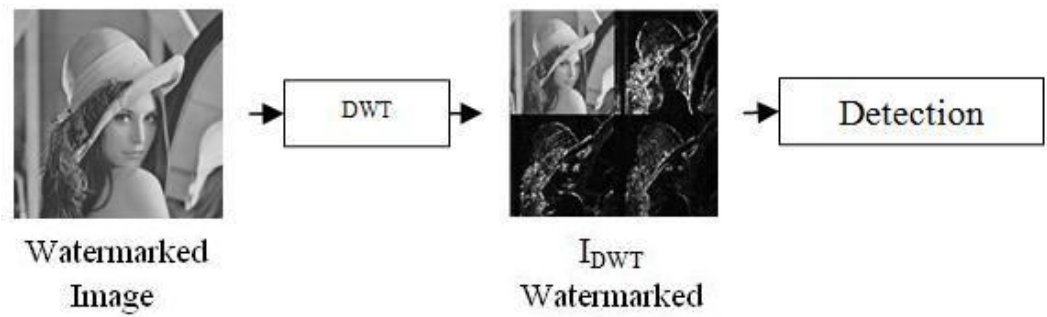

Figure 5. Extraction scheme of the brand

The first step isolates all blocks of N × N HH band. The next is to compute the gradient of each block. The mark is detected by determining the sign of sum of the gradient coefficients. The following algorithm summarizes the extraction process :

1. Compute the wavelet transform of the watermarked image.
2. Compute the gradient of each blocHH block (there GHH this gradient)

```
    If (Sum_Of_GHH_Coefficients > 0) then
          Bit_Of_The_Mark = 1
    Else
          If (Sum_Of_GHH_Coefficients < 0)
                Bit_Of_The_Mark = 0
          End If
    End If
```

## 4. RESULTS

The method was tested with "lena, mandrill and plane" images that have 512 × 512 pixels. The mark contains 1024 bits. The size of the block to be inserted is 8 × 8. We made several tests by changing the value of λ and λ = 20 gave an interesting result with a good compromise robustness / imperceptibility.

The following figures show respectively the original image (a), the inserted mark (b), the watermarked image (c) and the difference (d) between the original and the watermarked images :

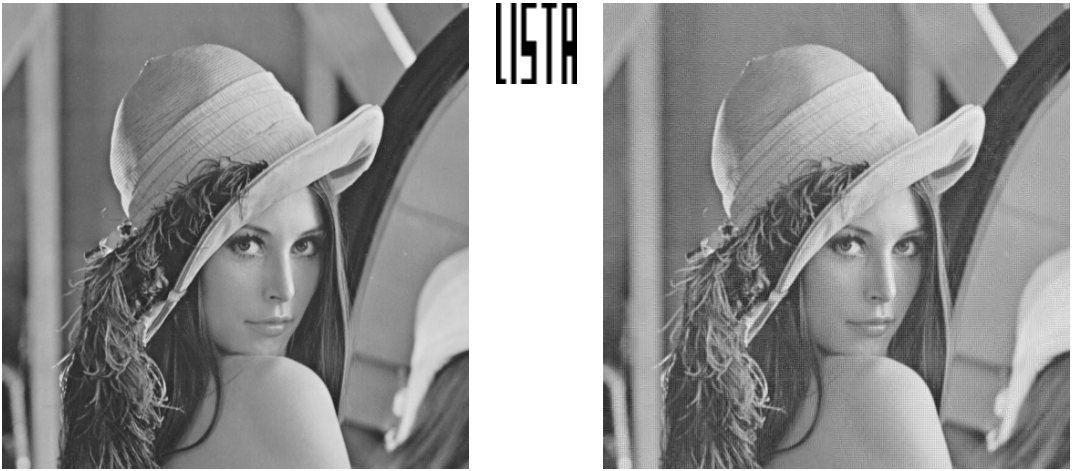

(a)       (b)       (c)

International Journal on Cryptography and Information Security (IJCIS), Vol.3, No. 3, September 2013

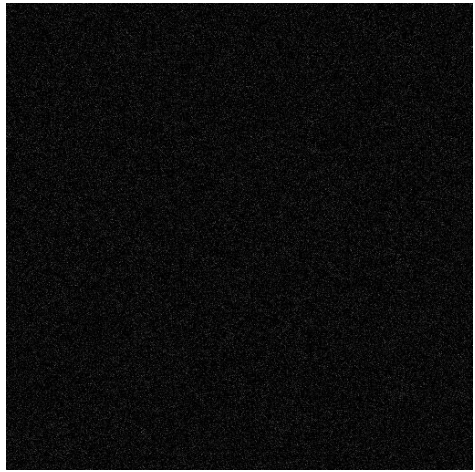

(d)

Figure 6. Original image (a), brand (b) watermarked image (c) difference (d)

The watermark is imperceptible because the average PSNR measured between the two images (Original and Watermarked) is 26 dB. Figure (d) also illustrates the invisibility of the mark.

The robustness of the technique was evaluated by attacking the watermarked image with filtering (median filter), addition of Gaussian noise and salt-pepper (different values of the noise standard deviation σ are shown in Table 1), JPEG compression (Quality Factor QF%), GIF color reducing and rotation (low rotation). The following table shows the extracted marks and means of obtained Bit Error Rate (BER) :

Table 1. Reconstructed marks for each type of attacks

| *Noise (Salt and pepper 0.05)* | *Noise (Salt and pepper 0.1)* | *Noise (Salt and pepper 0.2)* |
|---|---|---|
| BER = 0.0283 | BER = 0.0740 | BER = 0.2021 |
| *Noise (Gaussien 0.01)* | *Noise (Gaussien 0.05)* | *Noise (Gaussien 0.1)* |
| BER = 0.0088 | BER = 0.1152 | BER = 0.2012 |
| *Median Filtrer (3x3)* | *Median Filtrer (5x5)* | *Median Filtrer (7x7)* |
| BER = 0.0430 | BER = 0.3154 | BER = 0.2961 |
| *JPEG (QF 10%)* | *JPEG (QF 50%)* | *JPEG (QF 90%)* |

International Journal on Cryptography and Information Security (IJCIS), Vol.3, No. 3, September 2013

| 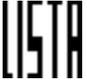<br>BER = 0 | 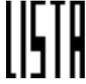<br>BER = 0 | 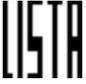<br>BER = 0 |
|---|---|---|
| *GIF Compression* | *Histogram Equalization* | *Rotation 0.25/-.025* |
| 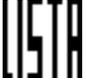<br>BER = 0 | 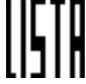<br>BER = 0 | 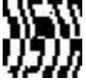<br>BER = 0.4414 |

## 5. CONCLUSION AND DISCUSSION

In this paper, we propose a new watermarking method "robust and blind" based on edge insertion. The image of Table 1 shows that our technique resists to several types of attacks : GIF and JPEG compression, noise addition, histogram equalization and rotation. Robustness against rotation is very limited, indeed to an angle greater than 0.25, the method is not more robust. Compared to the Kuraz [9] method and those described by Kumsawat [8] and Fang [10], ours, which gives a BER = 0 for all quality factors, provides better robustness to JPEG compression. As described by Oliveria et al. [15] "an algorithm that is resistant to all kinds of attacks does not exist", our algorithm can't solve all the problems in image watermarking field because it also has its limits (unrobustness with high rotation). But, the method can be improved by adjusting the technical parameters such as $\lambda$ or size blocHH :

- By increasing $\lambda$, we can improve the robustness of the method, but the imperceptibility of the mark is no longer respected.
- Capacity (number of bits that can be inserted in the host image) of our technique can be improved by reducing the size of blocHH. For example, the capacity could be doubled by halving the number of rows in the block, but this operation would reduce greatly the robustness of the method.The Sharkas et al. [16] method is also a good solution by inserting a portion of the mark in another part which will be added to the image host.

## ACKNOWLEDGEMENTS

The authors would like to thank IST-D (Institut Supérieur de Technologie d'Antsiranana) for its Sponsor and Financial Support.

International Journal on Cryptography and Information Security (IJCIS), Vol.3, No. 3, September 2013

## Authors

Henri Bruno Razafindradina was born in Fianarantsoa, Madagascar, on 1978. He received, respectively, his M.S degree and Ph.D in Computer Science and Information Engineering in 2005 and 2008. He served since 2010 as a lecturer at the Higher Institute of Technology Diego Suarez, became an assistant reviewer in 2011. His current research interests include images compression, multimedia, computer vision, information hiding.

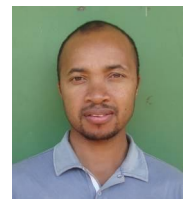

Karim Attoumani Mohamed is the Technical Assistant in Telecoms and Responsible of Procurement Management at the Project Implementation Unit (ABGE) for Comoros. He was the Head of the Department of Studies and Projects at the Comoros ICT Regulation Authority. He is interested in Image Processing, ICT and Internet governance

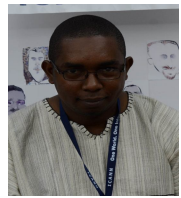